\documentclass[twocolumn,amsmath,amssymb]{revtex4}
\usepackage{graphicx}
\usepackage{color}
\usepackage[utf8]{inputenc}
\usepackage[T1]{fontenc}
\usepackage{soul}
\usepackage{epstopdf}

\begin{document}
\parindent=0pt

\title{\bf {The excitation function for Li+HF$\to$LiF+H at collision energies below 80\,meV}}

\author{Rolf Bobbenkamp\footnotemark{}\footnotetext{Present address: Carl Zeiss Optronics GmbH, 35576 Wetzlar, Germany}, Hansj\"urgen Loesch}
\affiliation{Fakult\"at f\"ur Physik, Universit\"at Bielefeld, 33501 Bielefeld, Germany}

\author{Marcel Mudrich, Frank Stienkemeier}
\affiliation{Physikalisches Institut, Universit\"at Freiburg, 79104 Freiburg, Germany}

\begin{abstract}
We have measured the dependence of the relative integral cross section of the reaction Li+HF$\to$LiF+H on the collision energy using crossed molecular beams. By varying the intersection angle of the beams from 37$^o$ to 90$^o$ we covered the energy range 25\,meV$\leq E_{\rm tr} \leq 131\,$meV. We observe a monotonous rise of the cross section with decreasing energy over the entire energy range indicating that a possible translational energy threshold to the reaction is significantly smaller than 25\,meV. The steep rise is quantitatively recovered by a Langevin-type excitation function based on a vanishing threshold and a mean interaction potential energy $\propto R^{-2.5}$ where $R$ is the distance between the reactants. To date all threshold energies deduced from \textit{ab-initio} potentials and zero-point vibrational energies are at variance with our results, however, our findings support recent quantum scattering calculations that predict significant product formation at collision energies far below these theoretical thresholds.
\end{abstract}


\maketitle


\section{Introduction}
The rapid development of techniques for cooling, trapping and manipulating cold and ultracold
molecules opens the opportunity of studying chemical reactivity in the low temperature
regime~\cite{Balakrishnan:2004,Krems:2008,Bell:2009,Hutson:2010}. Cold reactive collisions
are not only important limitations to stable trapping of molecules but also present
fascinating aspects of their own interest. In the cold collision regime, reactions are
governed by quantum dynamics involving tunneling and scattering
resonances~\cite{Krems:2008,Weck:2005}. For our study of cold collision phenomena we have
selected the title reaction due to both the availability of theoretical results and the applicability of a variety of experimental methods.

Attractive from the theoretical point of view is the small number of electrons together with
three chemically very different but light atomic constituents that make the slightly exoergic
($\approx$0.16\,meV) reaction Li+HF$\to$LiF+H an ideal prototype system for developing
methods to calculate reliable ab-initio potential energy surfaces
(PESs)~\cite{Balint-Kurti:1977,Zeiri:1978,Shapiro:1979,Chen:1980,Carter:1980,Palmieri:1989,
Parker:1995,Aguado:1997,Jasper:2002,Bobbenkamp:2005,Aguado_unpublished}  and to calculate
quantum scattering
phenomena~\cite{Parker:1995,Walker:1981,Lagana:1988,Baer:1989,Lagana:1991,Baer:1994,
BaerJCP:1994,Balint-Kurti:1993,Gogtas:1996,Aguado:1997,Lara:1993,Lara:1998,Lagana:1993,
Zhu:1997,Lagana:2000,LaganaCPL:2000,Hobel:2004,Weck:2005,Zanchet:2009}. Particularly
challenging for a quantitative prediction of observables is the low energy range near
threshold where quantum effects dominate and a precise knowledge of the PES around the
transition state is essential. Crucial for the assessment of the preciseness and reliability
of the applied theoretical methods is the comparison of experimental and computational
results. Very qualified for this purpose are certainly angular and velocity distributions of
products~\cite{Bobbenkamp:2005,HobelDiss} but also the dependence of the relative integral
reaction cross section (IRCS) on the translational collision energy, $E_{\rm tr}$, the
excitation function~\cite{BaerJCP:1994,Lara:1998,LaganaCPL:2000,Qiu:2006,Hobel:2001}. At
low collision energies the excitation function bears direct information on the existence and
size of a translational threshold that allows a sensitive examination of the shape of the PES
near the transitions state and the height of a possible potential energy barrier.

An inspection of computational results shows that all PESs
~\cite{Balint-Kurti:1977,Zeiri:1978,Shapiro:1979,Chen:1980,Carter:1980,Palmieri:1989,Parker:1995,Aguado:1997,Jasper:2002,Bobbenkamp:2005,Aguado_unpublished}
known to date feature a significant barrier at a bent transition state with substantial
height. However including zero point energies of both the reactants and the transition state
the semi-classical translational threshold energies deduced from these barrier heights turn
out to be markedly smaller. For example, for the most frequently used recent PESs, the
semi-classical thresholds and barriers amount to $\approx$27 and 182\,meV~\cite{Parker:1995},
$\approx$ 68\,meV and 233\,meV~\cite{Aguado:1997}, $\approx$56 and
221\,meV~\cite{Bobbenkamp:2005,Aguado_unpublished}, respectively. Neglecting quantum
scattering phenomena the semi-classical thresholds manifest the lowest collision energies
required for product formation from ground state reactants. An observation of products at
energies below these calculated thresholds could be a clue to marked quantum phenomena or to
insufficient accuracy of the PESs or to a combination of both. Recently, it has been
demonstrated in time-dependent wave packet scattering calculations
~\cite{Weck:2005,Zanchet:2009} based on the PES of
ref.~\cite{Bobbenkamp:2005,Aguado_unpublished} that resonances play an important role for
Li+HF. They lead to product formation at a significant rate at collision energies far below
the semi-classical threshold.\

One interesting experimental aspect of the title reaction is the availability of methods that
allow to investigate scattering phenomena in a wide range of collision energies from hot
(several 100\,meV) to cold (20\,meV) and ultra-cold ($<1$\,meV). In an earlier experimental study we
have measured the excitation function within the range 82\,meV $\leq E_{\rm tr}\leq$ 376\,meV
~\cite{Hobel:2001} using molecular beams intersecting at a 90$^o$ angle. The energy was varied
by applying the seeded beam technique. At energies above $\approx$ 120\,meV the function turns
out to be roughly constant but below it rises steeply with decreasing $E_{\rm tr}$ and assumes
at the lowest energies the largest values within the entire range. Consequently, a threshold
to reaction, if existing at all, has to be located below 82\,meV.

In the present paper we report the results of another crossed beam study on the excitation
function designed to lower the range of translational energy to 25\,meV. Variation of the
collision energy is achieved by varying the intersection angle of the beams (see below). We
find that in the extended energy range the IRCS continues to rise monotonously with decreasing $E_{\rm tr}$ thus shifting the upper boundary of a threshold to below 25\,meV.

It should be noted that a novel experimental setup using new techniques for preparing samples
of cold atoms and molecules is about to be completed. It will allow to continue our search for
the threshold of this process at energies below 1\,meV.

\section{Experimental Methods}

\begin{figure}[ht]
\includegraphics[width=\columnwidth,origin=c]{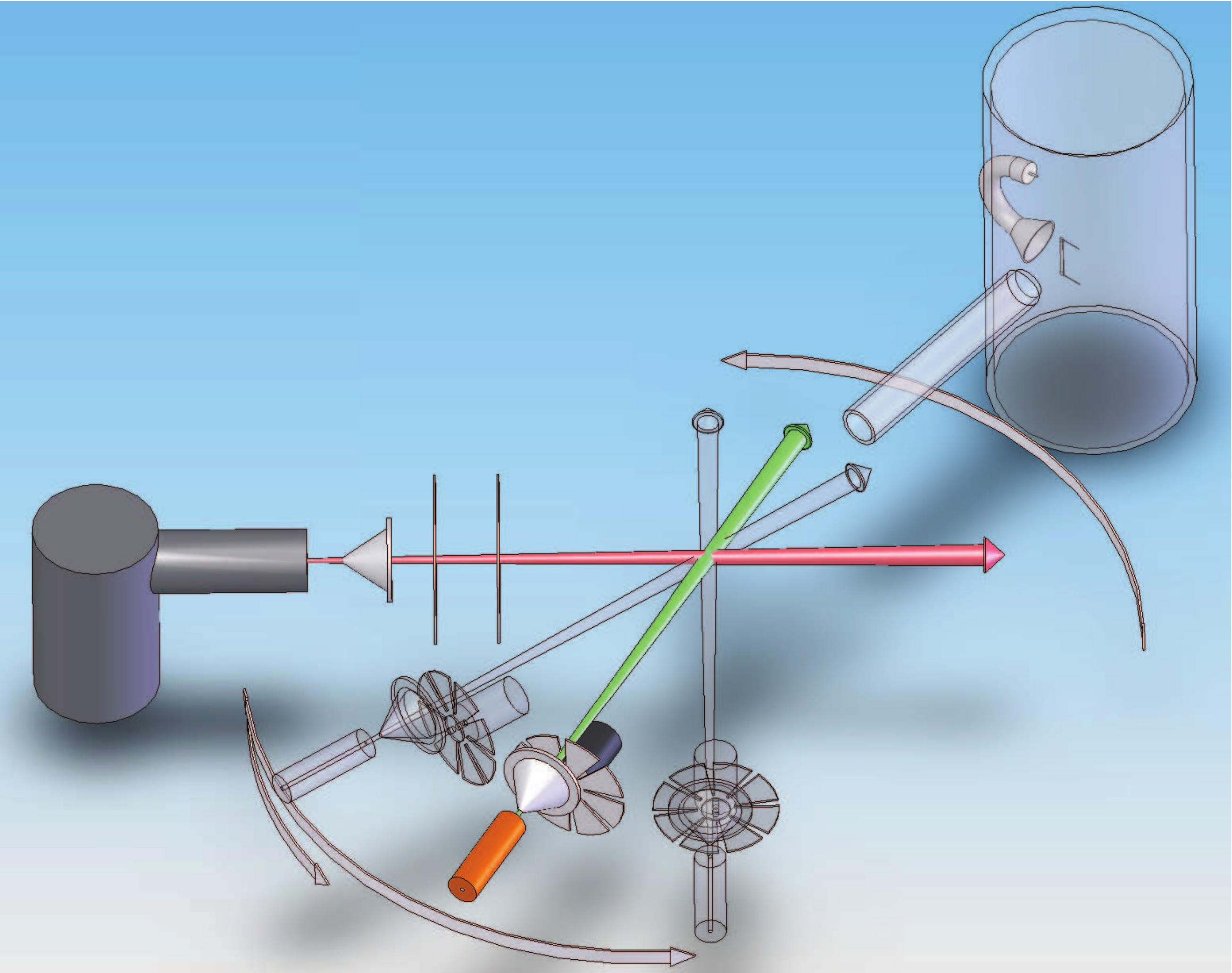}\\
\caption{Schematic view of the molecular beams arrangement. The setup is basically the same as
used in our previous experiments~\cite{Hobel:2001} except for the HF beam that can now be
rotated around the intersection volume to vary the intersection angle $\Gamma$ in the range
$37^o \leq \Gamma \leq 90^o$. An additional HF-beam modulator placed between nozzle and
skimmer is not shown for clarity.\label{fig:setup}}
\end{figure}
To achieve translational energies below 82\,meV (the lowest energy accessible in
ref.~\cite{Hobel:2001}) we abandon the usual 90$^o$ geometry and cross the beams at a variable
intersection angle~\cite{Buntin:1987,Hsu:1997,Naulin:1999}. Fig.~\ref{fig:setup} shows a
sketch of the molecular beams arrangement. It deviates from the one used in earlier
experimental studies~\cite{HobelDiss,Hobel:2001,Loesch:1993} only by the setup of the HF-beam.
The nozzle is now mounted within a differentially pumped vacuum chamber (not shown) that can
be rotated around the scattering volume. In this way the intersection angle $\Gamma$ of the
two beams and thus also the mean relative velocity
\begin{equation}
\bar v=[v_{\rm Li}^2 +v_{\rm HF}^2-2 v_{\rm Li}v_{\rm HF}\cos\Gamma]^{0.5}
\label{eq:relv}
\end{equation}
and the translational collision energy
\begin{equation}
E_{\rm tr}(\Gamma)=\frac{1}{2}\mu\bar v^2
\end{equation}
can be varied continuously. $\mu$ denotes the reduced mass of the colliding reactants and
$v_{\rm Li}, ~ v_{\rm HF}$ are their nominal velocity. Depending on $\Gamma$, significantly
smaller or larger collision energies can be achieved compared to the one obtained for the
usual perpendicular intersection.  In this study the two vacuum chambers housing the beam
sources are designed such that $\Gamma$ can be varied in the range $\rm 37^o\leq \Gamma \leq
90^o$. \bigskip

The experimental method and data acquisition follows closely the ones discussed in
 ref.~\cite{Hobel:2001}. Briefly, the neat HF beam is created by a nozzle heated to
 around 500\,K to avoid clustering. Stagnation pressure (250\,mbar) and nozzle
 temperature are always kept constant. The elevated temperature leads to a mean rotational energy
 of 8.9\,meV~\cite{Hobel:2001} and a population
 of the first vibrationally excited state of $<0.25\%$. The beam is chopped with a frequency of
 a few Hz using
 a tuning-fork like modulator or a chopper wheel at a duty cycle of 50\% and the intensity
 is monitored by a quadrupole mass spectrometer detector equipped with an electron
 bombardment ion source. The Li beam is diluted with Ne and its intensity is measured
 by surface ionization on a hot rhenium (Re) ribbon. During an experimental run only the
 intersection angle $\Gamma$ is varied while the operational conditions of the beams are kept
 constant. \bigskip

\begin{table}
\begin{center}
\caption{Beam velocity parameters, velocities in m/s}
\label{tab1} \vspace*{0.5cm}
\begin{tabular}{lcccc}
\hline
Run      & $u_{\rm Li}$ & $\alpha_{\rm Li}$     & $u_{\rm HF}$& $\alpha_{\rm HF}$ \\
\hline
A/B      &1530          & 221              & 1205        & 185 \\
C        &1750          & 290              & 1215        &195 \\
\end{tabular}
\end{center}
\end{table}
The velocity of both beams is determined by conventional time-of-flight (TOF) arrangements.
The (density) velocity distributions, $n(v)$, are extracted from the measured TOF profiles by
fitting the parameters $u,~ \alpha$ of the expression
\begin{equation}
n(v)= const~ v^2~ \exp\{-[(v-u)/\alpha]^2\}
\end{equation}
\noindent to the data. The parameters $u$ and $\alpha$ of the various data sets are compiled
in Table 1. The velocity spreads of the Li and HF beams $\alpha$ result in an experimental
relative uncertainty of the collision energy $E_{tr}$ of about 25\%. \bigskip

The products created in the intersection volume of the beams are detected via surface
ionization on a hot Re ribbon mounted in a separately pumped ultra high vacuum chamber
(residual gas pressure below $10^{-8}\,$mbar). A channeltron converts the ions desorbing from
the Re surface to electron pulses which are counted by a two channel scaler synchronized with
the HF beam modulator. One scaler counts the signal and background, the other only the
background pulses. The detected scattering intensity (signal) $I_{\rm tot}$ is then derived as
difference of the two scaler contents. Angular distributions of the signal $I_{\rm
tot}(\Theta)$ are obtained by rotating the main detector automatically around the intersection
volume in the plane of the beams within a wide range of laboratory (LAB) scattering angles
$\Theta$. Crucial for the data analysis (see below) is the knowledge of time-of-flight (TOF)
distributions of scattered particles at various scattering angles. They are measured employing
a fast spinning chopper wheel with 8 equally spaced slots (2\,mm wide) mounted between the
skimmer of the HF beam and the scattering volume (see Fig.~\ref{fig:setup}). The length of the
flight path is 254\,mm.\bigskip

\begin{figure}[ht]
\includegraphics[width=0.9\columnwidth,origin=c]{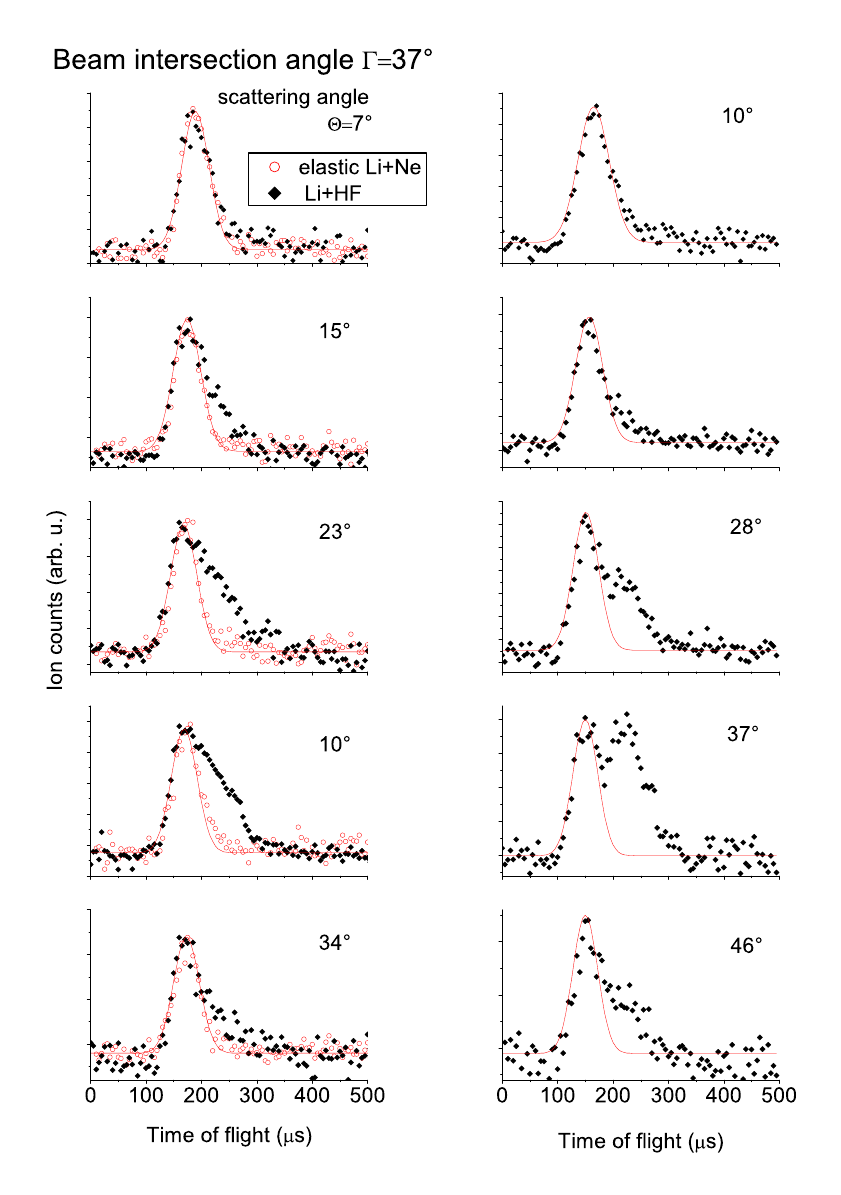}\\
\caption{Typical TOF profiles measured at $\Gamma=37^o$ and $50^o$ for the indicated
scattering angles. The profiles are used to separate elastic (smooth curves) from the detected
scattering intensity (data points). See text for more details.\label{fig:tof}}
\end{figure}
The surface ionization is not specific with respect to alkali atoms or alkali compounds and
detects both species with roughly the same efficiency. Therefore, the detected scattering
intensity $I_{\rm tot}$ is proportional to the sum of fluxes of both the elastically scattered
Li atoms $I_{\rm Li}$ and the products LiF, $I_{\rm LiF}$. To separate both components we
employ the TOF distributions; examples are displayed in Fig.~\ref{fig:tof} for $\Gamma=37^o$
and $50^o$. All TOF profiles feature a peak at short flight times and a broad shoulder or even
a second peak at longer times. At $\Gamma=90^o$ and higher Li velocities the peaks are always
well separated~\cite{Hobel:2001}.

\begin{figure}[ht]
\includegraphics[width=0.5\columnwidth,origin=c]{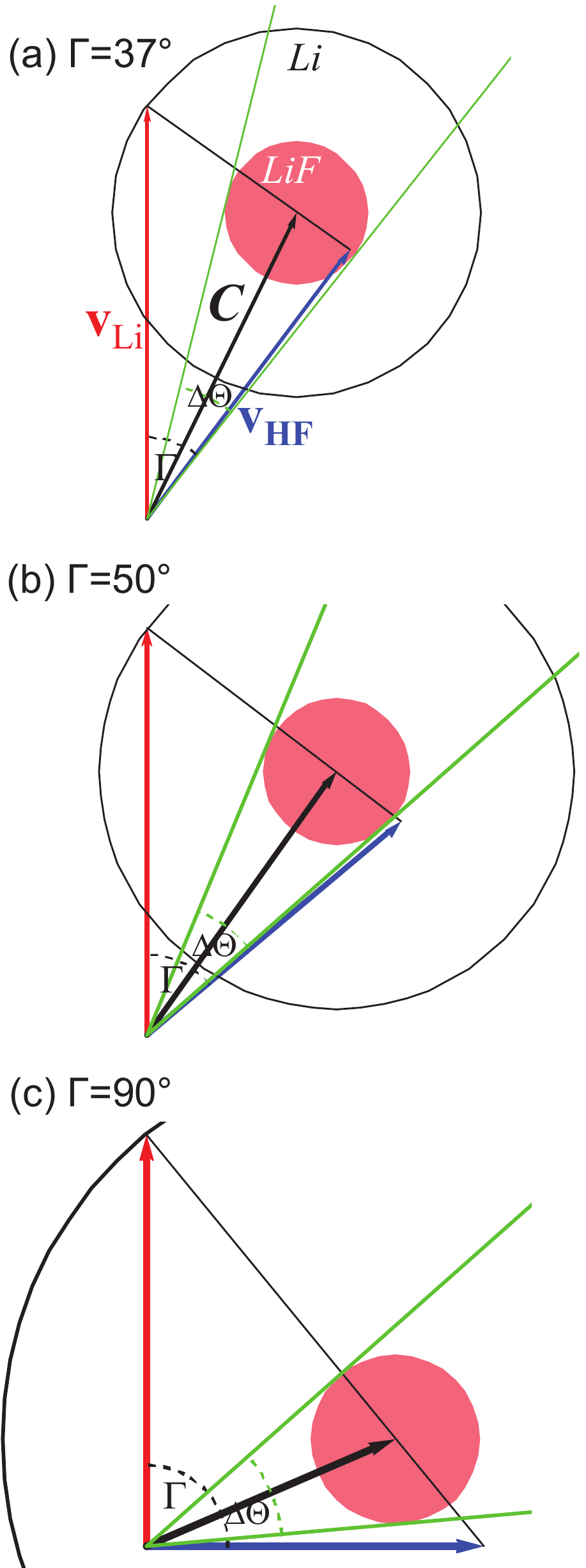}\\
\caption{Newton diagrams illustrating the kinematics for three beam intersection angles. The
radius of the outer circle around the tip of the centroid velocity vector $\bf C$ corresponds
to the velocity of the elastically scattered Li atoms in the CM system, the radius of the
inner filled circle to the maximal product velocity.\label{fig:newton}}
\end{figure}
The collision kinematics are illustrated in Newton-diagrams (Fig.~\ref{fig:newton}) relating
the LAB and center-of-mass (CM) frame velocities of reactants and products for beam
intersection angles $\Gamma=37^o,~50^o$, and $90^o$. The light arrows represent the laboratory
frame velocity vectors ${\bf v}_{\rm Li}$ and ${\bf v}_{\rm HF}$ of the reactants Li and HF
which include the intersection angle $\Gamma$. The radius of the outer circle around the tip
of the centroid velocity vector $\bf C$ corresponds to the velocity of the elastically
scattered Li atoms in the CM system, the radius of the inner shaded circle represents the
maximal product velocity. The light straight lines that are tangent to the inner circle
indicate the range of scattering angles $\Delta\Theta$ where reaction products LiF are
expected to occur in addition to elastically scattered Li atoms. Elastically scattered Li
atoms along $\bf C$ have higher speeds than the LiF products and will therefore reach the
detector at shorter flight times. Li atoms that are elastically scattered in the opposite
direction in the CM frame have low speeds in the LAB frame and will spatially disperse on
their way to the detector such that no significant contribution to the signal is expected.

\begin{figure}[ht]
\includegraphics[width=\columnwidth,origin=c]{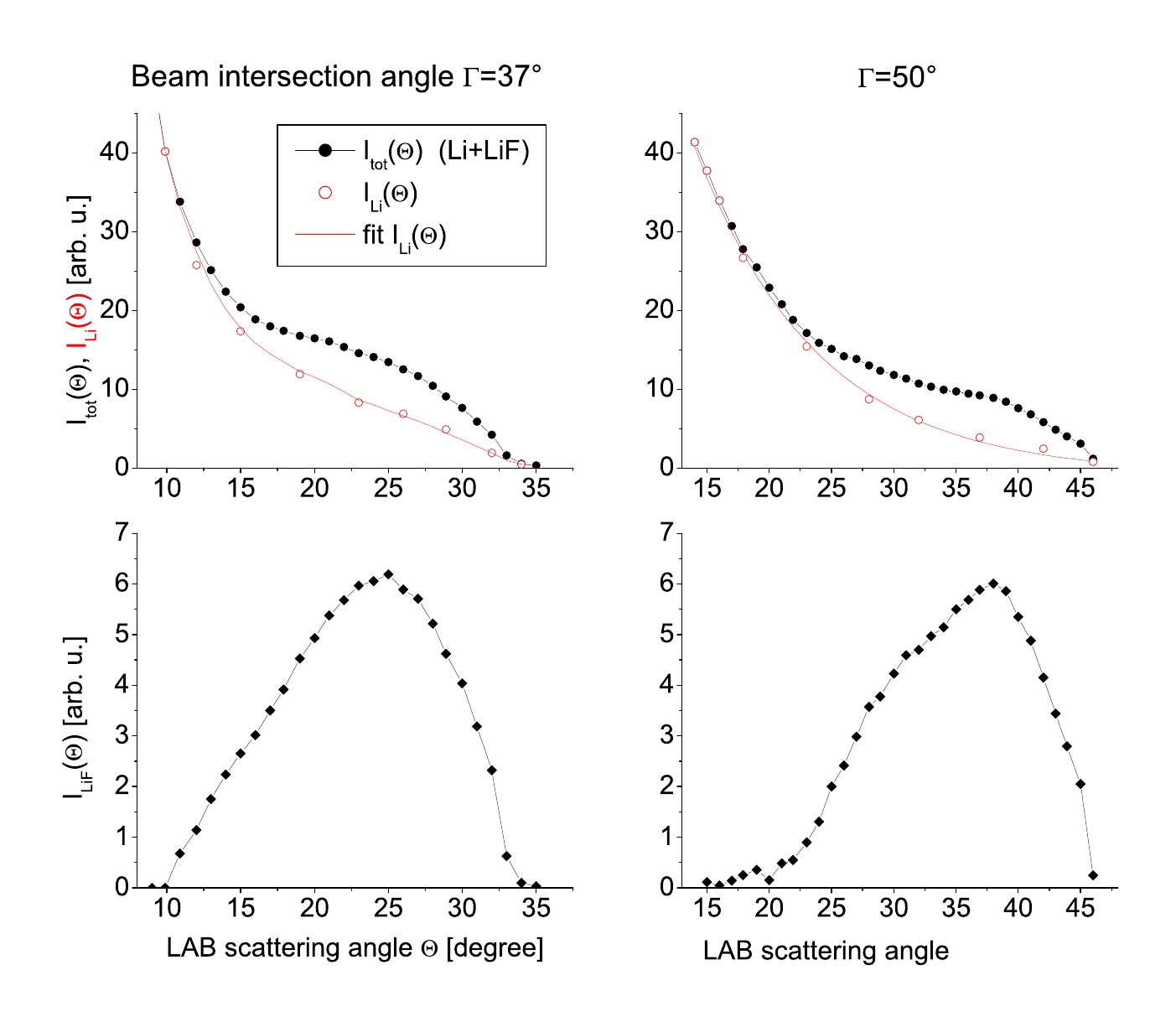}\\
\caption{Angular distributions of the detected (upper diagrams) and of the product
(lower diagrams) scattering intensity measured at $\Gamma=37^o$ and $50^o$. The open circles
in the upper diagrams mark the elastic intensity deduced from the TOF distributions of Figure
2. The solid line is the best fit interpolation used to deduce the reactive scattering
intensity (lower diagrams) by subtracting the elastic from the detected scattering
intensity.\label{fig:angledistr}}
\end{figure}
Thus, in Fig.~\ref{fig:tof} the peak at short flight times (fast particles) is attributed to
elastically scattered Li atoms while the one at later times (slow particles) is associated
with reaction products LiF. The solid line represents a numerically simulated TOF-distribution
of elastically scattered Li atoms whose reliability was confirmed by a comparison with TOF
profiles measured for the pure elastic system Li+Ne that features practically the same
kinematics as Li+HF (open circles in Fig.~\ref{fig:tof}). The elastic TOF profile is then
fitted to the fast slope of the data. The area below this scaled elastic peak over the area
below the entire TOF profile eventually provides the fraction of the detected scattering
intensity associated with elastically scattered atoms. For a given $\Gamma$ this fraction has
been determined at scattering angles for which TOF distributions are available and used to
isolate the angular distribution of the elastic component $I_{\rm Li}(\Theta)$ from the
detected scattering intensity $I_{\rm tot}(\Theta)$. The result of this procedure is shown in
Fig.~\ref{fig:angledistr} (upper panel). The open circles represent the angular distribution
of the elastic component $I_{\rm Li}(\Theta)$ and the solid line is a best fit curve through
these points. The difference between the detected scattering intensity and the elastic
component eventually yields the angular distribution of the product flux $I_{\rm
LiF}(\Theta)$ (lower panel in Fig.~\ref{fig:angledistr}). For more details
see~\cite{Hobel:2001}.

In principle the signal contribution identified with reactive scattering could also be
generated by rotationally inelastic collisions. At $\Gamma = 50^o$ inelastic collisions
Li+HF$(j=0)\rightarrow$Li+HF$(j=5)$ would yield Li atoms with similar CM speeds as the LiF
products. At $\Gamma = 37^o$, Li atoms could be generated at speeds similar to those of LiF in
inelastic collisions Li+HF$(j=0)\rightarrow$Li+HF$(j=3)$. Such selective rotation-changing
collisions appear quite implausible, though, given the common scaling laws that predict a fast
decay of the rotational inelastic collision cross section with increasing level
spacing~\cite{Raghavan:1985}.

\section{Results}

The IRCS, $\rm \sigma_r(E_{\rm tr})$, is proportional to the total flux of products $\dot
N_{\rm LiF}^{\rm total}$ generated in the scattering volume ${\cal V}$ and defined by the
expression
\begin{equation}
\sigma_r(E_{\rm tr})=\dot N_{\rm LiF}^{\rm total}/(n_{\rm Li}~n_{\rm HF}~\bar v~{\cal V}).
\label{eq:ircs}
\end{equation}
Here, $n_{\rm Li},~ n_{\rm HF}$ and $\bar v$ denote the number densities of the indicated
beams at the intersection volume and the mean relative velocity, respectively. Deviating from
our earlier study~\cite{Hobel:2001}, we leave the operational conditions of the beam sources
constant and vary only the intersection angle $\Gamma$. Thus both densities $n_{\rm HF}$ and
$n_{\rm Li}$ are constant, $\bar v$ can be easily deduced from the most probable beam
velocities (eq.~\ref{eq:relv}), and an inspection of the intersection geometry shows that
${\cal V} ={\cal V}_{90^o} /\sin \Gamma$ holds approximately.

The crucial quantity $\dot N_{\rm LiF}^{\rm total}$ is not directly accessible in the present
in-plane scattering experiment but can be deduced from the measured total in-plane product
flux $I_{\rm LiF}^{\rm in-plane}$ using the formal expression
\begin{equation}
\dot N_{\rm LiF}^{\rm total}={\dot N_{\rm LiF}^{\rm total} \over I_{\rm LiF}^{\rm in-plane}}
  I_{\rm LiF}^{\rm in-plane}.
\label{eq:formalflux}
\end{equation}
Inserting eq.~\ref{eq:formalflux} into eq.~\ref{eq:ircs} and suppressing all constant quantities we obtain the expression
\begin{equation}
\sigma_r(E_{\rm tr})\propto{\dot N_{\rm LiF}^{\rm total} \over I_{\rm LiF}^{\rm in-plane}}
 ~\sin\Gamma / \bar v ~ I_{\rm LiF}^{\rm in-plane}
 \label{eq:inplaneflux}
\end{equation}
relating the measured quantities and the relative integral reaction cross section. The ratio
of fluxes in eqs.~\ref{eq:formalflux} and \ref{eq:inplaneflux} corrects for the fraction of products that miss the detector; it can
be readily deduced from the relative differential reaction cross section (DRCS) in the
center-of-mass frame. In a previous study~\cite{HobelDiss} using perpendicularly intersecting beams we
have measured the relative DRCS at 6 energies within the range 82\,meV$\leq E_{\rm tr} \leq$
376\,meV and found that the ratio is constant with respect to the energy within an error
margin of $\pm 4\%$~\cite{Hobel:2001}. Unfortunately, a comparable extensive investigation of DRCSs is not yet available for the present low energy range. However, a preliminary analysis of the product
angular distributions measured at $\Gamma=37^o$ (24\,meV) and 50$^o$ (45\,meV) indicates that from
119\,meV~\cite{HobelDiss} ($\Gamma=90^o)$ to 24\,meV ($\Gamma=37{^o}$) a transition from the
forward/backward to a preferred sideways type DRCS occurs. Taking this into account we find
for all data sets that the product of factors left to the in-plane flux in eq.~\ref{eq:inplaneflux} is constant within the band width $\pm 5\%$. In view of a forthcoming more sophisticated determination of
the DRCS we suppress these small corrections and derive the relative IRCS from the data using
the simplified relation
\begin{equation}
\sigma_r(E_{\rm tr})\propto I_{\rm LiF}^{\rm in-plane}.
\label{eq:sigmainplane}
\end{equation}

We have measured three sets of angular distributions A, B, and C at various intersection
angles for the beam parameters given in Table 1. The constancy of the operational conditions
was checked carefully by measuring a reference angular distribution repeatedly during one run.
The intersection angles range from 37$^o$ to 90$^o$ (A,B) and 48$^o \leq \Gamma \leq 90^o$ (C)
corresponding to the energy range 25\,meV $\leq E_{\rm tr} \leq$ 108\,meV (A,B) and 50\,meV $\leq
E_{\rm tr} \leq$ 131\,meV (C). For each angular distribution of a given set we measured between
four and six TOF-profiles to separate elastic and reactive scattering.

\begin{figure}[ht]
\includegraphics[width=0.8\columnwidth,origin=c]{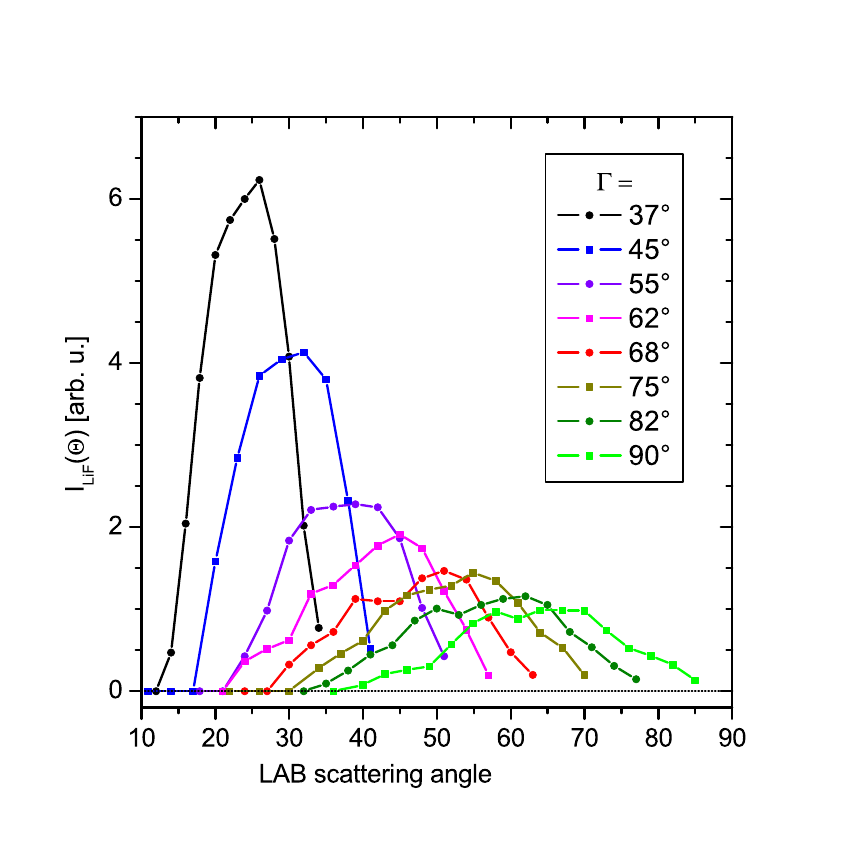}\\
\caption{Product angular distributions of data set A measured at the intersection angles
90$^o$ (open square), 82$^o$ (open circle), 75$^o$ (open triangle), 68$^o$ (open diamond,)
62$^o$ (solid square), 55$^o$ (solid circle), 45$^o$ (solid triangle) and 37$^o$ (solid
diamonds). Note the shift of the curves to smaller angles with decreasing $\Gamma$ caused by
the kinematics (Fig.~\ref{fig:newton}) and the strong increase of the peak intensity. The area
below the curves is the total product intensity used to determine the
IRCS.\label{fig:inplane}}
\end{figure}
As an example we show in Fig.~\ref{fig:inplane} the product angular distributions of set A.
With decreasing $\Gamma$ the curves shift to smaller LAB angles according to the changing
kinematics (Fig.~\ref{fig:newton}) and their peak intensities rise dramatically. The total
in-plane intensity $I_{\rm LiF}^{\rm in-plane}$ or $\sigma_r(E_{\rm tr})$
(eq.~\ref{eq:sigmainplane}) is given by the sum over all product intensities multiplied by the
Lab-angle increment (area below the curves),
\begin{equation}
I_{\rm LiF}^{\rm in-plane}=\sum I_{\rm LiF}(\Theta_{\rm i}) \Delta \Theta_{\rm i}.
\end{equation}

\begin{figure}[ht]
\includegraphics[width=0.8\columnwidth,origin=c]{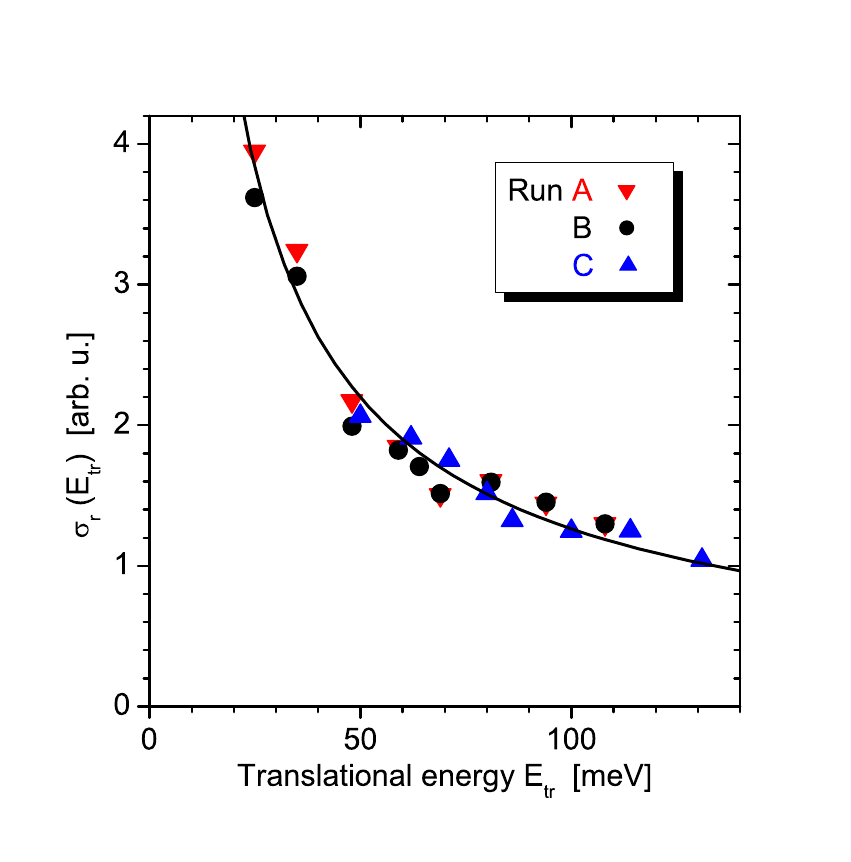}\\
\caption{The excitation function deduced from data set A (circles), B (triangles) and C
(diamonds). The data sets are mutually normalized to obtain the best agreement within the
overlap. The solid line through the data points is given by the simple power law
$\sigma_r(E_{\rm tr})\propto( 1/E_{\rm tr})^{0.8}$.\label{fig:sigmapowerfit}}
\end{figure}
\begin{table}
\begin{center}
\caption{Total in-plane scattering intensities in arbitrary units} \label{tab2}
\vspace*{0.5cm}
\begin{tabular}{lcccc}
\hline\\
$E_{\rm tr}$/meV  &$\Gamma$/degree  &        &$I_{\rm LiF}^{\rm in-plane}$&   \\
                  &                 &  A     &   B      &  C \\
\hline
 25               & 37              & 83411  &  110104  &  \\
 35               & 45              & 68510  &   93050  &  \\
 48               & 55              & 45930  &   60628  &  \\
 50               & 48              &        &          & 395200\\
 59               & 62              & 39040  &   55431  &   \\
 64               & 65              &        &   51909  &   \\
 69               & 68              & 31774  &   46050  &  \\
 71               & 60              &        &          & 335300\\
 80               & 65              &        &          & 289750\\
 81               & 75              & 31774   &   48590  &   \\
 86               & 68              &        &          & 253200\\
 94               & 82              & 30517  &   44170  &   \\
 100              & 75              &        &          & 238650\\
 108              & 90              & 27465  &   39540  &  \\
 114              & 82              &        &          & 239200\\
 131              & 90              &        &          & 199300\\
\end{tabular}
\end{center}
\end{table}
The results for set A are compiled in Table 2 together with those for B and C and displayed in
Fig.~\ref{fig:sigmapowerfit} as a function of the collision energy. The data points are
normalized such that they agree optimally within the overlapping energy range. Their
statistical error is not included in Fig.~\ref{fig:sigmapowerfit} for clarity. It can be
estimated from the scatter of the points and amounts to about $\pm 5\%$ essentially due to
uncertainties occurring in the process of separating elastic and reactive scattering. The
solid line through the data points represents the simple power law $\sigma_r(E_{\rm
tr})\propto( 1/E_{\rm tr})^{0.8}$.

\section{Discussion}

A strong motivation for performing the present experiments was the search for an answer to the
question raised in the earlier study~\cite{Hobel:2001}: How does the excitation function
continue below a collision energy of 82\,meV? The previous results indicated two
possibilities: either the excitation function assumes a maximum followed by a decline towards
a threshold as one would expect for a reaction featuring a non-vanishing translational
threshold energy or it continues to increase as expected for a reaction without threshold. The
results of the present investigation strongly support the assumption that no threshold hinders
the reaction of Li with HF. The excitation function continues to ascend monotonously at least
down to 25\,meV and is likely to continue this way~\cite{ArSeededLiBeam}. Due to the elevated
nozzle temperature of 500\,K, required to suppress dimerisation, low rotational states of HF
are populated with a mean rotational energy of 8.9\,meV  corresponding to a mean rotational
quantum number of $\bar j=1.4$ ~\cite{Hobel:2001}. Thus the indicated threshold energies refer
on average to the reaction Li+HF ($v=0$,$\bar j$=1.4). This internal energy may increase the
threshold of the ground state reaction Li+HF($v=0$,$j=0$) on average by the mean rotational
energy.

\begin{figure}[ht]
\includegraphics[width=\columnwidth,origin=c]{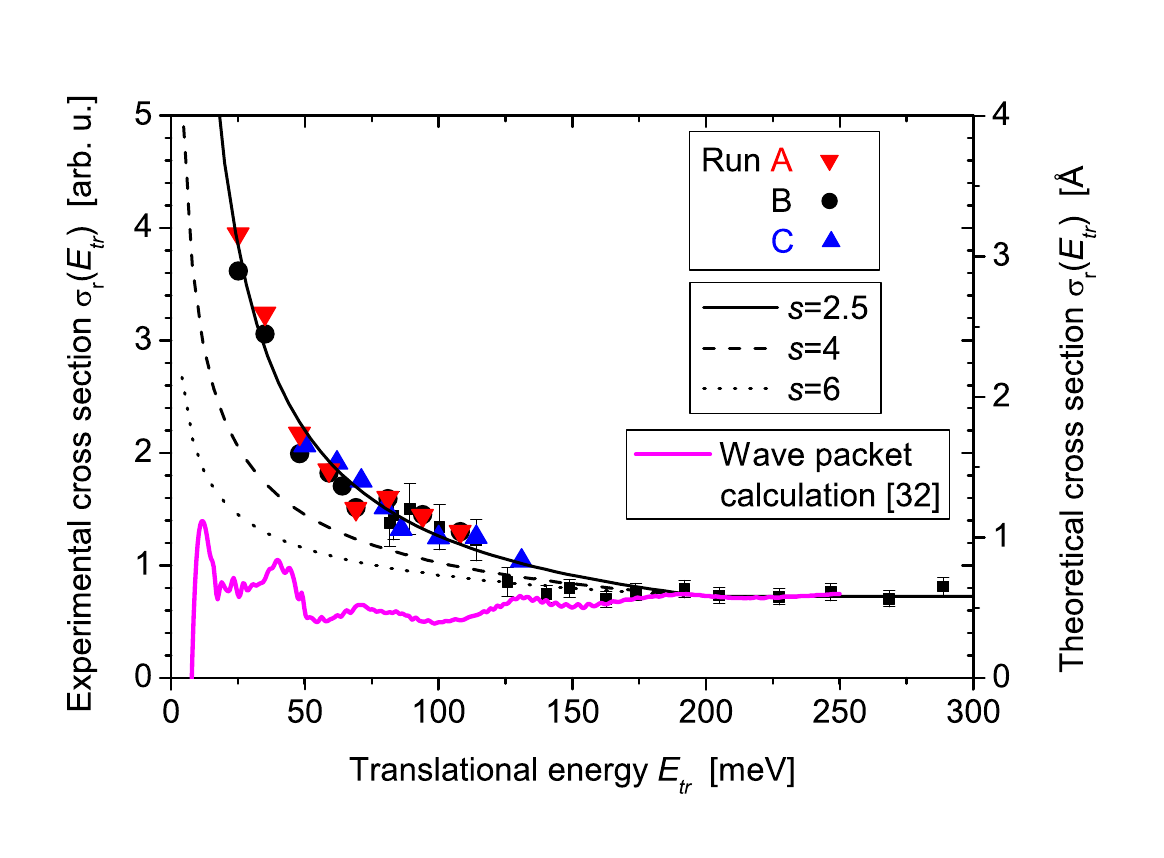}\\
\caption{Excitation function including the results of our previous study~\cite{Hobel:2001}
(solid squares). The new and earlier results are mutually normalized to obtain the best
agreement within the overlap. The curves at energies $\leq$ 200\,meV are Langevin cross
sections based on potential energy functions $\propto R^{-6}$ (dotted), $\propto R^{-4}$
(dashed) and $\propto R^{-2.5}$ (solid). The solid smooth line for energies $\geq 200$\,meV
refers to a constant hard sphere collision model matched to the Langevin functions at
200\,meV. The light solid line represents the result of recent wave packet
calculations~\cite{Zanchet:2009}.\label{fig:sigmapowerconst}}
\end{figure}
In Fig.~\ref{fig:sigmapowerconst} we have combined the results of the present and earlier
experiments and obtain an excitation function that covers now a wide energy range of more than
one order of magnitude from 25\,meV to 376\,meV. The curve can be subdivided into three
sections: above 200\,meV it is more or less constant, below 130\,meV the function rises
monotonously with decreasing $E_{\rm tr}$, and in between it gradually changes the shape
correspondingly. This behavior suggests the following classical reaction model.

(a) At energies $>200$\,meV  a rigid sphere collision mechanism~\cite{Qiu:2006,Levine} prevails where
the reaction occurs with a constant, energy independent probability $\cal P$ whenever the
impact parameter $b$ is smaller than the maximal one $b_{\rm max}$. The resulting constant
IRCS is then $\sigma_{\rm r}={\cal P} \pi b_{\rm max}^2$. The mechanism neglects long range
forces and thus all trajectories of the approaching reagents are straight lines.

(b) Below 130\,meV long range forces start getting important. The force curbs the trajectories
of the colliding reagents toward the center of the force and thus collisions with impact
parameters larger than $b_{\rm max}$ may react. With decreasing energy the influence of the
forces grows and thus the maximal impact parameter and the IRCS increase. Provided there is a
vanishing threshold the IRCS continues to rise and eventually diverges classically for $E_{\rm
tr}\to 0$. Assuming that (i) the interaction potential energy is $\propto 1/R^s$ where $R$ is
the distance between the reagents, and (ii) the reaction occurs with a constant probability
whenever the reagents overcome the effective potential's maximum, then the IRCS is described
by the "Langevin" power law $\sigma_{\rm r}(E_{\rm tr})\propto (1/E_{\rm
tr})^{2/s}$~\cite{Qiu:2006,Levine}. The best fit curve in Fig.~\ref{fig:sigmapowerfit} corresponds accordingly to an
interaction potential $\propto 1/R^{2.5}$. In Fig.~\ref{fig:sigmapowerconst} the $s=2.5$
result (solid line) is compared with curves referring to $s=6$ (dotted line) and $s=4$ (dashed
line) for energies below 200\,meV. The corresponding potentials describe the long range forces
of an atom-molecule ($s=6$) and of an ion-molecule system ($s=4$). Both curves feature the
typical steep rise but are clearly at variance with our data. The best fit power $s=2.5$
indicates that in our low energy range the chemical forces near the hard sphere radius are
responsible for the rise of the IRCS rather than the asymptotic long range forces with $s=6$.

(c) Between 130 and 200\,meV the transition between the hard sphere and the potential energy
dominated mechanism occurs.

The prediction of this classical model is illustrated in Fig.~\ref{fig:sigmapowerconst} as
solid line where we have omitted the transition region and matched directly the hard sphere
and Langevin functions at 200\,meV. The curve recovers the data nearly quantitatively over
this wide energy range.

The above classical model suggests a vanishing threshold but the experimental results provide
only an upper boundary and a small threshold may well be in agreement with our findings.
According to classical mechanics such a translational energy threshold is tightly related to
the height of the potential energy barrier $V_{\rm b}$ separating reagents and products.
Provided the approaching molecules are in their vibrational and rotational ground states
($v=0$, $j=0$) the formation of products requires that the translational energy of the
approaching particles exceeds $V_{\rm b}$. A non vanishing reaction threshold is then the
consequence of an existing barrier.

Quantum mechanics relaxes the tight classical relation between threshold and barrier height.
The colliding particles possess zero-point vibrational energy that varies during the approach
adiabatically from the zero-point energy of the free molecule $E^0_{\rm vib}$ to the one of the
triatomic aggregate at the barrier $E_{\rm b}^0$. Taking this into account, product formation
for ground state molecules ($v=0$,$j=0$) is then allowed in a classical sense, if the sum of
the collision and zero-point energy of the reagent molecule exceeds the sum of the potential
and zero-point energy at the barrier. The resulting semi-classical threshold energy
\begin{equation}
E_{\rm th}^0=V_{\rm b}+E_{\rm b}^0-E^0_{\rm vib}
\label{eq:eth}
\end{equation}
includes in addition to the barrier height $V_{\rm b}$ also the relevant zero-point energies $E_{\rm b}^0$ and $E^0_{\rm vib}$. However, in
contrast to classical mechanics, eq.~\ref{eq:eth} constitutes no rigorous lower limit for the
translational energy leading to product formation. Due to quantum phenomena such as tunneling
and resonances products may be formed also at collision energies far below this semi-classical
threshold (see below).

To date all PESs available for Li+HF predict a significant barrier height $V_{\rm
b}$~\cite{Balint-Kurti:1977,Zeiri:1978,Shapiro:1979,Chen:1980,
Carter:1980,Palmieri:1989,Parker:1995,Aguado:1997,Jasper:2002,
Bobbenkamp:2005,Aguado_unpublished}. Including the zero point energy of HF (256\,meV) and of
the transition state (95$\pm 10$\,meV depending on the geometric structure of the PES) the
semi-classical threshold (eq.~\ref{eq:eth}) for the most recent PESs amounts to 27\,meV
~\cite{Parker:1995}, 56\,meV~\cite{Bobbenkamp:2005,Aguado_unpublished} and
68\,meV~\cite{Aguado:1997}. The present experiments provide as upper boundary for the
threshold energy  25\,meV and, considering the significant product flux observed for the Ar
seeded Li experiment~\cite{ArSeededLiBeam}, 17\,meV. Furthermore, the excellent fit of the
data predicted by a model based on a vanishing threshold suggests a threshold located at
energies $\ll 17$\,meV. Thus our results suggest that the experimental threshold energy lies
markedly below the predicted semi-classical values. This discrepancy could be due to
insufficient accuracy of the computational methods used to calculate the PESs or to quantum
effects that promote reactions at energies far below the semi-classical threshold. The latter
is supported by recent wave packet calculations based on the PES of
~\cite{Bobbenkamp:2005,Aguado_unpublished} that predict a threshold for the total reaction
cross section at about 10\,meV which is qualitatively reconcilable with our
measurements~\cite{Zanchet:2009}. The theoretical excitation function~\cite{Zanchet:2009} is
displayed in Fig.~\ref{fig:sigmapowerconst}. The curve exhibits only a slight modulation
around the value $\sigma_r\approx 0.5\,$\AA\,\, as the translational energy is reduced down
from $E_{tr}=250\,$meV to a threshold value of about 10\,meV (see
Fig.~\ref{fig:sigmapowerconst}). The shape of the predicted excitation function departs
significantly from the experimental one but the threshold energy is in accord with the present
result. As the main source for product formation near threshold collisions with small $J$
(total angular momentum) have been identified. The reaction probabilities for $J=0$ exhibit a
rich spectrum of oscillations and resonances as a function of
energy~\cite{Weck:2005,Zanchet:2009} but these structures disappear if realistic IRCSs are
calculated by summing over all $J$.

\section{Conclusion}
The benchmark reaction Li+HF$\to$LiF+H has been studied at translational energies down to
25\,meV using a new crossed-beam apparatus with variable scattering angle between Li and HF
beams. This arrangement allows to tune the translational energy while keeping the beam source
conditions constant. The integral reactive scattering rate (excitation function) is deduced
from angle-resolved scattering as well as from time-of-flight traces in comparison with purely
elastic Li+Ne scattering. The resulting excitation function, which extends earlier
measurements to lower energies, steeply rises as the collision energy falls below about
150\,meV. This behavior is consistent with a barrier-less Langevin-type reactive process with
$R^{-2.5}$-scaling of the atom-molecule interaction potential. Alternatively, an energy
threshold below $\sim 25$\,meV may be present as predicted by recent wave packet
simulations~\cite{Zanchet:2009}.

Future efforts to further reduce the collision energy for conclusively disclosing or ruling out the existence of a reaction threshold will demand more sophisticated experimental approaches. To this end, a magneto-optical trap for preparing an ultracold Li scattering target is currently being set up and will be combined with a source for slow and cold molecules based on a rotating nozzle and electrostatic guiding~\cite{Strebel:2010}.

\begin{acknowledgments}
Support from the Deutsche Forschungsgemeinschaft is gratefully
acknowledged.
\end{acknowledgments}




\begin{thebibliography}{43}
\bibitem[{ArS()}]{ArSeededLiBeam}
\bibinfo{note}{In an experiment with an Ar seeded Li-beam we find at
  $\Gamma=37^o$ and $E_{\rm tr}=$ 17meV still a significant flux of products.
  However, we could not complete a full data set that overlaps with sets A and
  B.}
\end{thebibliography}

\end{document}